\documentclass[review]{elsarticle}

\usepackage{lineno,hyperref}
\usepackage{tikz-cd}
\usepackage{dsfont}
\usepackage{amsfonts}
\usepackage{amsmath}
\usepackage{slashed}
\usepackage{graphics}
\usepackage{amsmath}
\usepackage{amssymb}
\usepackage{tikz-cd}
\usepackage{setspace}
\usepackage{float}
\usepackage{cleveref}
\usepackage{units}
\usepackage{amscd}
\usepackage[left=1in, right=1in, top=1in, bottom=1in]{geometry}
\usepackage{nicefrac}
\modulolinenumbers[5]

\journal{IOP}
\begin{document}

\begin{frontmatter}

\title{\textbf{Neutrinos as possible probes for quantum gravity}}

\author{M.D.C. Torri$^{a*}$,}
\author{L.Miramonti$^{a}$}
\cortext[mycorrespondingauthor]{M.D.C.Torri: marco.torri@unimi.it, marco.torri@mi.infn.it}

\address[Unimi]{Dipartimento di Fisica, Universit\'a degli Studi di Milano and INFN Milano\\via Celoria 16, I - 20133 Milano, Italy}

\begin{abstract}
In this paper, we aim to explore the interplay between neutrinos and quantum gravity, illustrating some proposals about the use of these particles as probes for the supposed quantized structure of spacetime. The residual signatures of a more fundamental theory of quantum gravity can manifest themselves modifying the free particle dispersion relations and the connected velocity. In neutrino sector these supposed effects can modify the time of flight for astrophysical particles with different energies and can affect the usual neutrino oscillation pattern introducing species depending perturbations. We will highlight how perturbations caused by non-standard interactions in the solar neutrino sector can mimic the presumed quantum gravity effects. In fact, the mathematical formulation of non-standard interactions is equivalent to that of CPT-odd perturbations. We will, therefore, emphasize the need to identify the nature of different contributions in order to disentangle them in the search for quantum gravity effects. As a final point we will discuss the possibility to detect in the neutrino sector decoherence effects caused by the quantum gravity supposed perturbations. By reviewing current experimental constraints and observations, we seek to shed light on the intricate relationship between neutrinos and quantum gravity, and discuss the challenges and future directions in this fascinating field of research.
\end{abstract}


\end{frontmatter}

\setcounter{section}{0} 

\section{Introduction}
Symmetries are some of the most fundamental elements used in the formulation of every physical theory. Indeed, invariance properties give fundamental hints in deducing non-elementary particle structures and interaction laws. In this context, Lorentz invariance (LI) is the most important symmetry that underlies the formulation of Quantum Field Theory (QFT), that is the mathematical tool used in particle physics to describe interactions.\\
The quest for a comprehensive theory that unifies gravity with the principles of quantum mechanics remains one of the most significant challenges in modern physics. The search for candidate quantum gravity (QG) effects can be related to supposed departures from the fundamental LI symmetry as a signature of the foreseen quantum structure of spacetime \cite{COST}.\\
To investigate covariance modification, two primary approaches emerge: one involves the assumption of Lorentz invariance violation (LIV), exemplified by the Standard Model Extension (SME) \cite{SME}. The other entails modifications to Lorentz symmetry, as seen in Doubly Special Relativity (DSR) \cite{DSR} and the Homogeneously Modified Special Relativity (HMSR) theories \cite{HMSR}.\\
In the context of SME, LI is violated by the introduction of fixed background field justified by the string theory perspective. The foreseen modifications can be distinguished into two different families, introducing CPT-even \cite{SME} and CPT-odd \cite{SMECPT} perturbations to the usual formulation of the Standard Model of particle physics. The DSR and HMSR theories are motivated by a dynamical perspective and introduce modifications of the covariance formulation depending on the energy of the particle probing the spacetime. Additionally, the momentum space is envisioned to possess a non-trivial geometry designed to encode the proposed QG structure. Furthermore, within the framework of these models, Lorentz covariance is modified and acquires an explicit dependence on the non-trivial and curved momentum space geometry. The modified Lorentz transformations can be viewed as a particular class of diffeomorphisms relating the different momenta spaces associated to different observers.\\
From an experimental perspective, astrophysics represents one of the most promising sectors for the search of QG evidence, since the envisioned phenomenological effects will be more pronounced, benefiting from the high energies and long free propagation path of cosmic messengers \cite{Whitepaper,Coleman, Stecker, Sarkar,EllisGamma, Torri1, Torri2, Torri3}. Within the realm of astroparticles, neutrino can assume an important role. In this study we will investigate the potential for detecting residual QG signatures in neutrino sector, utilizing them as probes of the spacetime structure.\\
Since their discovery, neutrinos have captured interest for decades, playing a fundamental role in our understanding of the subatomic world. These neutral and weakly interacting particles have consistently provided opportunities for investigating new physics. The well-established oscillation phenomenon, described by the introduction of neutrino mass eigenstates, is a first example of physics beyond the Standard Model (BSM). Due to their weak interaction and lack of electric charge, neutrinos offer a means to explore physics across scales, from the atomic level to cosmological distances. Remarkably, the Universe is essentially transparent to their propagation. Moreover, unlike cosmic rays, neutrinos are not deflected by extragalactic magnetic fields, allowing them to be precisely aimed at their sources. Therefore, neutrinos exhibit the potential characteristics to serve as ideal probes for investigating gravity interactions at both the classical and quantum levels.\\
In the realm of neutrino physics, QG can leave its mark through modifications of the dispersion relations, affecting the neutrino propagation and resulting in energy depending particle velocity \cite{Amelino1}. Consequently non-identical time of flight for different energy neutrinos may be observed, as exemplified in multi-messenger investigation of GRB \cite{Amelino2,Ellis,Jacob,Wei,GuettaPiran} and supernova explosions \cite{Sigl,Moura}.\\
Neutrino sector can also be used to investigate the universality of the particle interplay with gravity. Supposing non universal QG perturbations, these modifications can affect the neutrino oscillation phenomenon, modifying the behaviour of neutrino transition between different flavor states during propagation. Indeed, by altering the dispersion relations, QG can introduce species-dependent propagation velocities, modifying the foreseen oscillation pattern \cite{Torri4,Torri5}. In the following sections we will illustrate how this phenomenon can potentially be detected in the context of atmospheric neutrinos.\\
When searching for QG effects, it is crucial to recognize that similar perturbations can arise within the framework of other new physics theories, such as neutrino non-standard interactions (NSI). NSI can influence the interplay between neutrino and matter beyond the predictions of the SM. Notably, we will illustrate that the SME mathematical description of CPT-odd Lorentz invariance violation (LIV) is identical to the NSI hamiltonian formulation \cite{Sarker,Torri6}. Understanding the origin and disentangling these various contributions, including NSI in matter interactions, becomes crucial for experimental analyses that aim to identify the presumed QG perturbations affecting the neutrino behavior \cite{Barenboim}. After that, we will proceed to analyze the possibility of identifying decoherence effects on neutrino propagation, caused by QG. \\
At the beginning, we will provide a brief review of how neutrinos can be employed to test scenarios related to various classical gravity theories, that is, non-quantum gravity models. Then, we introduce some of the best-known QG theoretical frameworks that enable phenomenological predictions. This introduction will encompass both covariance-modifying theories and the SME, which predicts LIV. We will illustrate how the possibility for non-universal corrections is introduced in all of these models. Subsequently, we will present the phenomenological effects that can be foreseen in the neutrino sector, potentially serving as QG signatures. We will explain how it is possible to search for these effects, disentangling them from other signatures caused by new physics phenomena, such as NSI. This approach can permit imposing constraints on the magnitude of induced perturbations or even to observe new phenomena. Finally we will consider some QG scenarios that can provide detectable decoherence effects on the propagation of neutrino beams.

\section{Neutrino and classical gravity}
Before focusing on the main topic of this work, the connection between neutrinos and QG, we present a brief review of the interplay between neutrinos and non-quantum gravitational theories. In this section, we will provide a brief overview of potential scenarios in which the neutrino sector can serve as a test bench for validating the predictions of the classical description of gravity, as obtained for instance in the framework of General Relativity (GR).\\
The presence of a gravitational field can generate a detectable effect within the neutrino sector, perturbing the oscillation pattern. Indeed, in the case of neutrino propagating in a curved spacetime, the gravitational interaction can induce a modification in the oscillation probability that can be detected in interference experiments \cite{Giunti}.\\
It is well known that the neutrino oscillation can be described including neutrino mass eigenstates in an extension of the Standard Model (SM) of particle physics. In the Hamiltonian picture the propagation is ruled by the Schr\"oedinger equation, written in the flavor basis as: $i\partial_{t}|\nu\rangle=H_{\alpha\beta}|\nu_{\beta}\rangle$. Considering the propagation from a point A to a point B, the solution of the previous equation is:
\begin{equation}
|\nu_{\alpha}(t)\rangle=\sum_{\beta}\sum_{k}U^{*}_{\alpha k}\exp\left(-i\int_{t_A}^{t_B}E_{k}\,dt+i\int_{\vec{r}_A}^{\vec{r}_B}\vec{p}\cdot d\vec{x}\right)U_{\beta k}|\nu_{\beta}(0)\rangle
\end{equation}
using the Pontecorvo-Maki-Nakagawa-Sakata (PMNS) unitary matrix $U$ to transform the flavor eigenstates (Greek indices) in mass eigenstates (Latin indices) and vice versa.
Assuming the neutrino as a relativistic particle, after travelling a distance $L$ in a flat spacetime, under the approximation $p_{i}\cdot x\simeq (E_{i}-p_{i})L$, the phase ruling the oscillation probability becomes:
\begin{equation}
\label{phase1}
\phi_{k}=(E_{k}-p_{k})L\simeq \frac{m_{k}^2}{2E}L
\end{equation}
and the transition amplitude can be written as a function of the propagation length as:
\begin{equation}
A_{\alpha\beta}=\langle\nu_{\beta}|\nu_{\alpha}(L)\rangle=\sum_{\beta}\sum_{k}U^{*}_{\beta k}e^{-i\phi_{k}L}U_{\alpha k}
\end{equation}
where the Greek index $\beta$ indicates the sum over the 3 flavor eigenstates and the Latin index $k$ stands for the sum over the corresponding mass eigenstates.\\
The oscillation probability is obtained as the squared of the transition amplitude:
\begin{equation}
P(\nu_{\alpha}\rightarrow\nu_{\beta})=|A_{\alpha\beta}|^2=|\langle\nu_{\beta}|\nu_{\alpha}(L)\rangle|^2.
\end{equation}
and it acquires the explicit form:
\begin{equation}
\label{prob}
P(\nu_{\alpha}\rightarrow\nu_{\beta})=\delta_{\alpha\beta}-4\sum_{j>k}\mathfrak{Re}\left(U_{\alpha j}U_{\beta j}^{*}U_{\alpha k}^{*}U_{\beta k}\sin^2{\left(\Delta\phi_{jk}\right)}\right)+2\sum_{j>k}\mathfrak{Im}\left(U_{\alpha j}U_{\beta j}^{*}U_{\alpha k}^{*}U_{\beta k}\sin^2{\left(\Delta\phi_{jk}\right)}\right)
\end{equation}
where $\Delta\phi_{ij}=\phi_{i}-\phi_{j}$ is the phase difference related to two different mass eigenstates.\\
In the presence of a gravitational field, in the GR framework, the propagation of a neutrino flavor eigenstate is:
\begin{equation}
U_{\alpha k}\exp\left(-i\int_{t_A}^{t_B}E_{k}\,dt+i\int_{\vec{r}_A}^{\vec{r}_B}\vec{p}\cdot d\vec{x}\right)|\nu_{k}\rangle=U_{\alpha k}\exp\left[-i\int_{r_A}^{r_B}\left(\eta_{\mu\nu}+\frac{1}{2}h_{\mu\nu}\right)p_{k}^{\,\mu}\,dx^{\nu}\right]|\nu_{k}\rangle
\end{equation}
where $h_{\mu\nu}=g_{\mu\nu}-\eta_{\mu\nu}$, $g_{\mu\nu}$ is the metric of the considered curved spacetime and $\eta_{\mu\nu}$ is the usual Minkowski metric.
As a consequence, the phase \cref{phase1} acquires an additional term caused by the gravity induced perturbation:
\begin{equation}
\phi_{ij}=\phi_{ij}^{(0)}+\phi_{ij}^{(GR)}(M).
\end{equation}
In the simplest case of a Schwarzschild spacetime \cite{Cardall,Ahluwalia}, with source mass M, the GR additional phase can be written as:
\begin{equation}
\phi_{ij}^{(GR)}(M)=\frac{m_{ij}^2L}{2E}\left[\frac{GM}{r_{B}}-\frac{GM}{L}\ln{\left(\frac{r_{B}}{r_{A}}\right)}\right]
\end{equation}
where $\Delta m_{ij}^{2}=m_{i}^2-m_{j}^2$ is the difference of the squared mass eigenvalues.\\
Moreover, the neutrino sector has been associated to the possibility of testing the weak equivalence principle in the context of classical  \cite{Krauss,Gasperini,Mann,Blasone} and quantum gravity (as discussed below). From a more theoretical perspective, neutrino physics has been proposed as a test bench for the predictions of torsion gravity models \cite{Hammond}. Furthermore, extended models of gravity \cite{Capozziello,LambiaseCS,Capozziello2,Atayde,Buoninfante} and exotic geometry frameworks can also be tested in the context of this physical sector.

\section{Quantum gravity theoretical introduction}
New physical effects are expected at energies lower to the Planck's scale as residual evidence of a more fundamental theory about the quantum structure of spacetime. The physics described by GR and that described by QFT emerge at extremely different energy scales. Nevertheless, there are continuing efforts to identify a quantized theory of gravity to address the challenges caused by the non-unified formulation of the fundamental physical laws of nature. Some theories have been proposed to obtain a consistent formulation of QG, such as String theory \cite{Green}, Loop Quantum Gravity (LQG) \cite{Rovelli} and the non-commutative geometry scenarios \cite{Aastrup}. So far, no direct experimental observations confirm the quantum nature of the gravitational interaction. Some models have been developed in order to foresee testable QG effects. In this work, we will review some of these phenomenological frameworks, unified by the idea that QG induced perturbations can affect LI, leading to modifications of the covariance formulation. The consequent dynamical alterations can manifest in the free propagation and interaction of particles.

\section{DSR theories}
One approach employed to explore the potential covariance modification involves the construction of a comprehensive physical theory aimed at reconciling the existence of a second universal constant, the Planck length, with the relativity principle, associated to the constant speed of light. In this approach, named \emph{Relative Locality} \cite{Amelino3}, the momentum space acquires the role of fundamental playground where interactions take place. The covariance of physical laws is preserved, but in an amended formulation, that is the Lorentz transformations are modified. The usual spacetime is considered only as a local projection of the more fundamental phase space and can emerge only from the local measurements made by observers.

\subsection{DSR geometry structure}
The phase space geometry is postulated to exhibit a non-trivial structure, encoded in the modification of the composition rule of momenta \cite{Amelino3}. The particular choice of composition rules, selected for this study, results associative, but non-commutative:
\begin{equation}
\label{composition}
(p\oplus q)_{0}=p_{0}+q_{0}\qquad (p\oplus q)_{j}=p_{j}+e^{l\,p_{0}}q_{j}
\end{equation}
where $l$ represents the inverse of the Planck mass $l=1/M_{P}$, that is the parameter characterizing this kind of theories.\\
Using the definition of the modified composition rules the geometry structure of phase space can be obtained, starting from the definition of the connection:
\begin{equation}
\label{christoffel}
\Gamma^{\mu\nu}_{\alpha}=-\frac{\partial}{\partial p_{\mu}}\frac{\partial}{\partial q_{\nu}}(p\oplus q)_{\alpha}\bigg|_{p=q=0}.
\end{equation}
This construction leads to the infinitesimal parallel transport relation:
\begin{equation}
(p\oplus dq)_{\mu}=p_{\mu}+\tau^{\alpha}_{\mu}dq_{\alpha}=p_{\mu}+dq_{\mu}-\Gamma^{\alpha\beta}_{\mu}p_{\alpha}dq_{\beta}+\dots
\end{equation}
The non-trivial structure of the phase space is encoded by the non-commutativity of the composition rules of momenta, that gives rise to the space torsion:
\begin{equation}
-T^{\mu\nu}_{\alpha}=\frac{\partial}{\partial p_{\mu}}\frac{\partial}{\partial q_{\nu}}\left(p\oplus q-q\oplus p\right)_{\alpha}\bigg|_{p=q=0}.
\end{equation}
Instead, the curvature, arising from the non-associativity of the composition rules, results in its identically null value:
\begin{equation}
R^{\mu\nu\alpha}_{\beta}=\frac{\partial}{\partial p_{[\mu}}\frac{\partial}{\partial q_{\nu]}}\frac{\partial}{\partial k_{\alpha}}\left(\left(p\oplus q\right)\oplus k-p\oplus\left(q\oplus k\right)\right)_{\beta}\bigg|_{p=q=k=0}=0.
\end{equation}
Therefore, the phase space has a flat but non-trivial geometry. This result is possible since the momentum space geometry presents non-null torsion, that results in a non-metric origin.\\
The kinematical symmetries of this space can be described in the framework of Hopf algebra and have the structure of the $\kappa$-Poincaré group. The generators of $\kappa$-Poincaré satisfy the commutation rules \cite{Majid,Lukierski,Lukierski2}:
\begin{align}
\label{algebra}
&[P_{\mu},P_{\nu}]=0,\qquad [R_{i},R_{j}]=i\,\epsilon_{ijk}R_{k},\qquad [N_{i},N_{j}]=-i\,\epsilon_{ijk}R_{k}, \notag\\
&[R_{i},P_{0}]=0,\qquad\, [R_{i},P_{j}]=i\,\epsilon_{ijk}P_{k},\qquad\; [R_{i},N_{j}]=-i\,\epsilon_{ijk}N_{k},\\
&[N_{i},P_{0}]=i\,P_{i},\quad [N_{i},P_{j}]=i\,\delta_{ij}\left(\frac{1}{2l}\left(1-e^{-2P_{0}l}\right)+\frac{l}{2}|\vec{P}|^2\right)-i\,lP_{i}P_{j} \notag
\end{align}
where $\{P_{\mu}\}$, $\{N_{j}\}$, $\{R_{j}\}$ are the translation, boost and rotation generators.\\
The explicit form of the Poincaré algebra generators can be obtained starting from the previous commutation rules. For instance, in the context of the non commutative $\kappa$-Minkowski spacetime \cite{Agostini}, characterized by the coordinate commutation rule:
\begin{equation}
[x^{\mu},\,x^{\nu}]=i\zeta^{\mu\nu}_{\alpha}x^{\alpha} \qquad
\left\{
\begin{array}{ll}
\zeta^{\mu\nu}_{\alpha}=1\;\;\text{for}\,\mu=0,\,\nu=\alpha\in\{1,\,2,\,3\}\\
\zeta^{\mu\nu}_{\alpha}=0\;\;\text{for all the other indices combinations}\\
\end{array}
\right.
\end{equation}
the translation $\{P_{\mu}\}$ and the rotation $\{R_{i}\}$ generators are not modified with respect to the usual ones:
\begin{equation}
R_{i}=\epsilon_{ijk}x^{j}P_{k}
\end{equation}
whereas the boost generators $\{N_{i}\}$ are amended:
\begin{equation}
N_{i}=x^{0}P_{i}+x^{i}\left(\frac{1}{2l}\left(1-e^{-2P_{0}l}\right)+\frac{l}{2}|\vec{P}|^2\right)-l\,x^{j}P_{i}P_{j}.
\end{equation}
The Hopf algebra bicross-product formalism associated to the $\kappa$-Poincaré group is given by:
\begin{align}
\label{bicross}
&\Delta P_{0}=P_{0}\otimes\mathbb{I}+\mathbb{I}\otimes P_{0},\qquad \Delta P_{i}=P_{i}\otimes \mathbb{I}+e^{-l\,P_{0}}\otimes P_{i},\\
&\Delta R_{i}=R_{i}\otimes\mathbb{I}+\mathbb{I}\otimes R_{i},\qquad \Delta N_{i}=N_{i}\otimes\mathbb{I}+e^{-l\,P_{0}}\otimes N_{i}-l\,\epsilon_{ijk}e^{l\,P_{0}}P_{j}\otimes R_{k}
\end{align}
with the coalgebra antipodes $S$ and counits $\epsilon$ associated to the generic generators $\{P_{\mu}\}$ given by:
\begin{align}
&S(P_{\mu})(p)=(\ominus p)_{\mu}\Rightarrow S(E)=-E,\quad S(P)=-e^{l\,E}P,\quad S(N)=-e^{-l\,E}N\\
&P_{\mu}(0)=\epsilon(P_{\mu})\Rightarrow \epsilon(E)=\epsilon(P)=\epsilon(N)=0
\end{align}
where $\ominus$ is the inversion related to the modified sum rules of momenta.\\
Starting from the defined commutation rules \cref{algebra}, the modified Lorentz transformations in the momentum space can be computed exponentializing the generators   \cite{Bruno,Amelinotrasf,Gubitosi}.\\
The line element acquires the explicit form of a de Sitter spacetime and it is invariant under the action of the modified transformations:
\begin{equation}
ds^2=dE^2-e^{2l\,E}dp^2.
\end{equation}
It is convenient to introduce the associated vierbein $e^{\mu}_{\,\nu}(p)=diag(1,e^{-l\,p_{0}},e^{-l\,p_{0}},e^{-l\,p_{0}})$, so that the line element can be written as:
\begin{equation}
ds^2=e^{\alpha}_{\,\mu}(p)e^{\beta}_{\,\nu}(p)\eta^{\mu\nu}dp_{\alpha}dp_{\beta}.
\end{equation}
Since the momentum modified composition rule covariance is not guaranteed by these transformations:
\begin{equation}
\Lambda(\xi,\,p\oplus q)\neq\Lambda(\xi,\,p)\oplus\Lambda(\xi,\,q),
\end{equation}
the backreaction must be introduced:
\begin{equation}
\Lambda(\xi,\,p\oplus q)=\Lambda(\xi,\,p)\oplus\Lambda(\xi\lhd p,\,q)
\end{equation}
where $\Lambda$ is the generalized Lorentz transformation, $\xi$ represents the rapidity and $p$ and $q$ are the momenta associated to the interacting particles. The backreaction is defined in relative locality as the transformation induced on the momentum of the second particle interacting, since the rapidity is changed in a momentum dependent way.\\
The structure of the $\kappa$-Poincaré algebra of generators can be obtained in the context of a de Sitter momentum space introducing the following projective coordinates \cite{Arzano}:
\begin{align}
P_{0}&=\frac{1}{l}\sinh{(l\,p_{0})}+l\frac{|\vec{p}|^2}{2}e^{l\,p_{0}},\notag\\
P_{j}&=p_{j}e^{l\,p_{0}},\notag\\
P_{4}&=\frac{1}{l}\cosh{(l\,p_{0})}-l\frac{|\vec{p}|^2}{2}e^{l\,p_{0}}.
\end{align}
In the deSitter basis the commutator rules are the ordinary Poincaré ones. On the other hand using the projective basis the non-linear commutator structure of the $\kappa$-Poincaré algebra \cref{algebra} can be obtained.\\
In DSR theories, a significant portion of the phenomenology arises from modifications to the kinematics, as encoded in the modified dispersion relations (MDR). Two different approaches are used, in this context, to defining the MDR, the first one consists of using the geodesic distance \cite{Amelino4}:
\begin{equation}
\label{mdr}
d(p,\,m)=\frac{1}{l}\text{arcosh}\left(\cosh{(l\,E)}-\frac{l}{2}e^{-l\,E}|\vec{p}|^2\right)=m^2.
\end{equation}
From the previous relation it is simple to obtain the following equality:
\begin{equation}
\label{mdr1}
m^2\simeq E^2-e^{l\,E}|\vec{p}|^2.
\end{equation}
In the other approach the MDR are defined using the algebra Casimir operator:
\begin{equation}
\label{mdr2}
\frac{4}{l^2}\sinh^2{(l\,E/2)}-e^{l\,E}|\vec{p}|^2\simeq m^2.
\end{equation}
In both cases the MDR given by \cref{mdr1,mdr2} can be approximated at the first perturbative order in the following form:
\begin{equation}
\label{mdrdef}
E^2-|\vec{p}|^2(1+l\,E)=m^2.
\end{equation}

\subsection{Construction of the spacetime coordinates}
Spacetime is an auxiliary idea constructed in the attempt of describing the interaction of different particles. Indeed, the coordinate space is a concept that emerges from the dynamic of different particles interaction in the momentum space.\\
The dynamic of interacting particles can be defined applying a variational principle to the action:
\begin{equation}
\label{action}
S=\sum_{j\in J}\int d\tau\bigg\{-x_{(j)}^{\mu}\dot{p}_{\mu}^{(j)}+\mathcal{N}_{j}C^{j}(p,\,m)+\zeta^{\mu}\mathcal{K}_{\mu}\left(p_{1}(\tau),\ldots,p_{n}(\tau)\right)\bigg\}
\end{equation}
where the index $j$ represents the j-esim particle. The MDR condition \cref{mdr,mdr1}:
\begin{equation}
C^{j}(p,\,m)=d^2(E,\,p)-m^2
\end{equation}
is enforced in the action using the Lagrange multipliers $\mathcal{N}_{j}$. Additionally, $\zeta^{\mu}$ are the multipliers used to impose the conservation rules $\mathcal{K}_{\mu}(p_{1},\ldots,p_{n})$. These conservation rules are constructed starting from the modified composition rules of momenta \cref{composition} using the $\oplus$ sum operation.\\
From the variation of the action \cref{action} the equation of motions can be computed:
\begin{align}
&\frac{dx^{\mu}_{(j)}}{dt}=-\mathcal{N}_{j}\frac{\partial d(p,\,m)}{\partial p_{\mu}^{(j)}},\notag\\
&\frac{dp_{\mu}^{(j)}}{dt}=0,
\end{align}
alongside the relations constraining the dispersion relations and the composition rules:
\begin{align}
&d^2(E,\,p)-m^2=0,\notag\\
&\mathcal{K}_{\mu}(p_{1},\ldots,p_{n})=0.
\end{align}
The coordinates of the emerging spacetime, related to the position of the j-esim particle, can be constructed employing the boundary terms of the action. The interaction coordinates can be computed as:
\begin{equation}
x^{\mu}_{j}(0)=\zeta^{\nu}\frac{\partial}{\partial p_{\mu}^{(j)}}\mathcal{K}_{\mu}(p_{1},\ldots,p_{n}).
\end{equation}
As a consequence, ignoring the momentum space curvature, i.e., at the leading order, all the interaction worldlines meet at a single spacetime point. The introduction of the momentum space curvature justifies the idea of \emph{Relative Locality} \cite{Amelino3}, indeed any translated observer detects the interactions at a distant set of events.\\
In the case of conservation rules $\mathcal{K}_{\mu}(p_{1},\ldots,p_{n})=p_{1}\oplus\ldots\oplus p_{n}$, the modified Lorentz transformations of the coordinates can be computed applying the covector diffeomorphism rule:
\begin{equation}
x'^{\mu}_{(j)}=x^{\nu}_{(j)}\frac{\partial p_{\nu}^{(j)}}{\partial p_{\mu}'^{(j)}}.
\end{equation}
The action of the translation on the coordinates can be computed applying the conservation rules of momenta \cite{Amelino3}:
\begin{equation}
\delta x^{\mu}_{(j)}=\epsilon_{\nu}\{\mathcal{K}_{\nu},\,x^{\mu}_{(j)}\}=\epsilon^{\mu}-\epsilon^{\nu}\,\Gamma_{\nu}^{\mu\alpha}p_{\alpha}^{(j)}.
\end{equation}
where the connection \cref{christoffel} has been used. Since the momentum space has a non-trivial geometry and the composition rules are not linear, the translations happen in an energy depending manner. This phenomenon is an explicit manifestation of the \emph{Relative Locality}.\\
As a final remark, it is possible to demonstrate the invariance under the action of the modified Lorentz transformations of all the quantities defined in the previous equations \cite{Gubitosi}.

\subsection{Non-universal scenario in DSR theory}
\label{nusDSR}
In this section we will illustrate, in the context of DSR theories, how QG perturbations of particle kinematics can acquire a non-universal character, by supposing the deformations of the kinematical symmetries depending on the particle species. In this context it is necessary to introduce a mathematical tool that can combine the modified algebras associated to the different particle species interacting.\\
We begin by introducing a method for combining the algebras and then, we introduce the associated mixed coproduct. Finally, we deduce how to add, that is to compose, the four-momenta related to different particle species within the Hopf algebra formalism. In this context the application of the mixing coproduct must be defined as the mathematical structure generating the modified composition rules of momenta belonging to different algebras.\\
For instance, the coproduct of the elements belonging to different algebras $H_{j}$ tagged by the index $j$ \cite{Amelino5,Amelino6} can be defined introducing a support algebra $H'$ associated with a projection map $\phi_{j}$:
\begin{equation}
\phi_{j}:H_{j}\rightarrow H'.
\end{equation}
In the following we furnish an explicit example on how the projection map can be defined in the case of $\kappa$-Poincaré algebras underlying the different particle symmetry groups. The map $\phi$ can be constructed relating the bicrossproduct basis generators of the different algebras in the following way:
\begin{equation}
\phi(P_{\mu})=\frac{l'}{l}P_{\mu}',\quad\phi(R_{i})=R'_{i},\quad\phi(N_{i})=N'_{i}
\end{equation}
where $l$ and $l''$ are the characteristic coefficients of the different $\kappa$-Poincaré algebras.\\
Using the previous definitions it is now straightforward to obtain the relations:
\begin{align}
&[\phi(P_{\mu}),\,\phi(P_{\nu})]=0=\phi([P_{\mu},\,P_{\nu}]),\notag\\
&[\phi(R_{i}),\,\phi(R_{j})]=i\,\epsilon_{ijk}R'_{k}=\phi([R_{i},\,R_{j}]),\notag\\
&[\phi(N_{i}),\,\phi(N_{j})]=-i\,\epsilon_{ijk}R'_{k}=\phi([N_{i},\,N_{j}]),\notag\\
&[\phi(R_{i}),\,\phi(N_{j})]=i\,\epsilon_{ijk}N'_{k}=\phi([N_{i},\,P_{j}]),\notag\\
&[\phi(R_{i}),\,\phi(P_{0})]=0=\phi([R_{i},\,P_{0}]).
\end{align}
Finally, supported by the previous results, it is possible to verify the following equations:
\begin{align}
&[\phi(R_{i}),\,\phi(P_{j})]=\frac{l'}{l}[R'_{i},\,P'_{j}]=i\,\epsilon_{ijk}\frac{l'}{l}P'_{k}=\phi([R_{i},\,P_{j}]),\notag\\
&[\phi(N_{i}),\,\phi(P_{0})]=i\,\frac{l'}{l}P'_{i}=\frac{l'}{l}[N'_{i},\,P'_{0}]=\phi([N_{i},\,P_{0}])
\end{align}
and finally the relation:
\begin{align}
&[\phi(N_{i}),\,\phi(P_{j})]=i\,\delta_{ij}\left(\frac{1-e^{-2l'P'_{0}}}{2 l}+l\left(\frac{l'}{l}\right)^2|\vec{P}'|^2\right)-i\,l\left(\frac{l'}{l}\right)^2P'_{i}P'_{j}=\\\notag
&=\frac{l'}{l}\left(i\,\delta_{ij}\left(\frac{1-e^{-2l'P'_{0}}}{2l'}+l'|\vec{P}'|^2\right)-i\,l' P'_{i}P'_{j}\right)=\phi([N_{i},\,P_{j}]).
\end{align}
As a consequence, demonstrating that $H'$ is a Hopf algebra, it is possible to state that the projection map preserves the commutation rules.\\
The mixed coproduct, associated with $H'$, can be defined composing the projection maps related to different Hopf algebras $H_{j}$, for instance, in the case of two different algebras:
\begin{align}
\label{mixcoprod}
&\qquad\Delta':H'\otimes H'=\phi_{1}(H_{1})\otimes\phi_{2}(H_{2})\notag\\
&\begin{CD}
H'@>\Delta'>>H'\otimes H'@>\phi_{1}^{\,-1}\otimes\phi_{2}^{\,-1}>>H_{1}\otimes H_{2}.
\end{CD}
\end{align}
In the coalgebra sector the following relations are valid for the coproduct:
\begin{align}
&\Delta'(P'_{0})=\Delta'(\phi(P_{0}))=\frac{l'}{l}P'_{0}\otimes\mathbb{I}+\mathbb{I}\otimes\frac{l'}{l}P'_{0}=\phi\otimes\phi(\Delta(P_{0})),\notag\\
&\Delta'(P'_{i})=\Delta'(\phi(P_{i}))=\frac{l'}{l}P'_{i}\otimes\mathbb{I}+e^{-l'P'_{0}}\mathbb{I}\otimes \frac{l'}{l}P'_{i},\notag\\
&\Delta'(R'_{i})=\Delta'(\phi(R_{i}))=R_{i}\otimes\mathbb{I}+\mathbb{I}\otimes R_{i},\notag\\
&\Delta'(N'_{i})=\Delta'(\phi(N_{i}))=N'_{i}\otimes\mathbb{I}+e^{-l'P'_{0}}\otimes N'_{i}+l'\epsilon_{ijk}P'_{j}\otimes R'_{k}.
\end{align}
A last check is needed to prove that the map $\phi$ is an isomorphism, proving the compatibility of the morphism $\phi$ with the antipodes and the counits, usually defined in the context of Hopf coalgebra. We verify the compatibility with the antipodes via the relations:
\begin{align}
&\phi(S(P_{0}))=-\frac{l'}{l}P'_{0}=S'(\phi(P_{0})),\notag\\
&\phi(S(P_{i}))=-\frac{l'}{l}e^{-l'P'_{0}}P'_{i}=S'(\phi(P_{i})),\notag\\
&\phi(S(R_{i}))=-R_{i}=S'(\phi(R_{i})),\notag\\
&\phi(S(N_{i}))=-e^{l'P'_{0}}N'_{i}+l'\epsilon_{ijk}e^{l'P'_{0}}P'_{j}R'_{k}=S'(\phi(N_{i})),
\end{align}
where $S'$ represents the antipode map related to the algebra $H'$. The compatibility with the counit follows straightforwardly. The definition of the inverse map $\phi^{-1}:H'\rightarrow H$ is simply obtained in the form:
\begin{equation}
\phi^{-1}(P'_{\mu})=\frac{l}{l'}P_{\mu},\quad\phi^{-1}(R'_{i})=R_{i},\quad\phi^{-1}(N'_{i})=N_{i}.
\end{equation}
Using the previous results it is possible to state that the map $\phi:H\rightarrow H'$ is an isomorphism connecting the $\kappa$-Poincaré algebras $H$ and $H'$.\\
Starting from the relation \cref{mixcoprod}, we can introduce the momentum modified composition rules associated with quantities belonging to different spaces, tagged with a superscript index:
\begin{align}
&p_{0}^{\,(1)}\oplus'_{l_{1}\,l_{2}}q_{0}^{\,(2)}=\frac{l'}{l_{1}}p_{0}^{\,(1)}+\frac{l'}{l_{2}}q_{0}^{\,(2)},\notag\\
&p_{i}^{\,(1)}\oplus'_{l_{1}\,l_{2}}q_{i}^{\,(2)}=\frac{l'}{l_{1}}p_{i}^{\,(1)}+\frac{l'}{l_{2}}e^{-l p_{0}^{\,(1)}}q_{i}^{\,(2)}.
\end{align}
In the previous equalities $p$ and $q$ are the momenta related to the two different particle species interacting, which probes the modified phase spaces characterized respectively by the parameters $l_{1}$ and $l_{2}$.\\
The inverted maps and the reverse order composition rules simply amounts to the exchange of the deformation coefficients $l_{1}$ and $l_{2}$.\\
Thanks to the illustrated results, it is possible to assert that within the framework of DSR theories, non-universal corrections can be admitted. In fact, the algebras related to the modified symmetries of different species of particles can coexist. This last achievement allows us to state that also in DSR scenario different particles can admit different MDR in the form of \cref{mdrdef} and with species-dependent corrections.

\section{HMSR theory}
Considering the momentum space as the realm where the particle interactions take place, the structure of spacetime can be obtained in the context of the Hamilton geometry. This is the case of the HMSR model \cite{HMSR}, constructed starting from the MDRs defined as homogeneous functions, hence ensuring their geometric origin in a generalized Finsler geometry framework. The MDR are supposed to present a particle species dependence and are defined as:
\begin{equation}
\label{Finslermdr}
F(\mathbf{p})=E^2-\left(1-f\left(\frac{|\vec{p}|}{E}\right)\right)=m^2.
\end{equation}
As a result the phase space acquires a non-trivial geometric structure that encodes the supposed QG perturbations.
The peculiar form of the perturbation function assures that in the high energy limit every particle presents a different maximum attainable velocity: $f\left(|\vec{p}|/E\right)\rightarrow\epsilon$, an important result that will be used subsequently.\\
The MDR can be considered as a norm defined in the phase space, thus presenting a geometric origin, compatible with the momentum metric:
\begin{equation}
\label{metric1}
g^{\mu\nu}(p)=\left(
                \begin{array}{cc}
                  1 & 0 \\
                  0 & -\mathbb{I}_{3\times3}(1-f(|\vec{p}|/E)) \\
                \end{array}
              \right).
\end{equation}
The coordinate space metric can be obtained as the inverse of the previous one using the relation:
\begin{equation}
g_{\mu\alpha}(x(p))g^{\alpha\nu}(p)=\delta_{\mu}^{\nu}.
\end{equation}
As already shown, it can be useful the introduction of the generalized vierbein in order to write the coordinate space metric as:
\begin{equation}
g_{\mu\nu}(x)=e^{\alpha}_{\,\mu}(p)\,\eta_{\alpha\beta}\,e^{\beta}_{\,\nu}(p)=
              \left(
                \begin{array}{cc}
                  1 & 0 \\
                  0 & -\mathbb{I}_{3\times3}(1+f(|\vec{p}|/E)) \\
                \end{array}
              \right).
\end{equation}
The vierbein can be used to project every local physical quantity on a common Minkowski spacetime used to set the different particle species interaction. As a first result the spacetime coordinates and the momenta acquire a locality dependence on the energy of the space probe considered: $\tilde{x}^{\mu}=e^{\mu}_{\,\nu}(p)x^{\nu}$ and $\tilde{p}_{\mu}=e_{\mu}^{\,\nu}(p)p_{\nu}$. As a consequence the coordinates do not commute anymore and the commutation rules are:
\begin{align}
\label{commutationsHMSR}
&\left[\tilde{x}^{\mu},\,\tilde{x}^{\nu}\right]=\theta^{\mu\nu}\notag\\
&\left[\tilde{p}_{\mu},\,\tilde{p}_{\nu}\right]=0\notag\\
&\left[\tilde{x}^{\mu},\,\tilde{p}_{\nu}\right]=\delta^{\mu}_{\,\nu}-\frac{\epsilon}{3}\left(\frac{|\vec{p}|}{E}\right)^{2}
\end{align}
where the antisymmetric matrix $\theta^{\mu\nu}$ is defined as:
\begin{align}
&\theta^{ij}=\theta^{00}=0\notag\\
&\theta^{i0}=-\theta^{0i}=\epsilon
\end{align}
where $\epsilon$ is a constant encoding the magnitude of the QG perturbation and defines the first order series expansion of the function $f\left(\frac{\vec{p}}{E}\right)\simeq\epsilon\cdot\left(\frac{\vec{p}}{E}\right)$. The commutation rules obtained here are similar to the definition of the non-commutative coordinates of \cite{Witten}.\\
The Clifford algebra, related to the geometry structure, is amended and can be defined using the vierbein projectors:
\begin{equation}
\Gamma^{\mu}=e^{\mu}_{\,\nu}(p)\gamma^{\nu}\;\Longrightarrow\;\{\Gamma^{\mu},\,\Gamma^{\nu}\}=2g^{\mu\nu}(p)
\end{equation}
and the Dirac equation can be written as:
\begin{equation}
\left(i\Gamma^{\mu}\partial_{\mu}-m\right)\psi=0.
\end{equation}
utilizing the modified Dirac matrices $\Gamma$.\\
Analogously, the generalized spinors can be defined and normalized, and finally, a minimal extension of the SM of particle physics can be obtained in this context.\\
As illustrated before for the DSR theories, also in the context of HMSR the covariance can be preserved. The generalized Lorentz transformations are obtained by projecting the standard ones on the support Minkowski spacetime utilizing the vierbein as projectors:
\begin{equation}
\Lambda_{\mu}^{\,\nu}(p)=e_{\mu}^{\,\alpha}(p)\,\Lambda_{\alpha}^{\,\beta}e_{\beta}^{\,\nu}(p)\;\Longrightarrow\;g_{\mu\nu}(\Lambda x)=\Lambda_{\mu}^{\,\alpha}(p)\,g_{\alpha\beta}(x)\,\Lambda_{\nu}^{\,\beta}(p).
\end{equation}
Therefore, the covariance is promoted to invariance under the action of a particular class of diffeomorphisms.\\
In HMSR model the particle kinematics is constructed starting from the action:
\begin{equation}
\mathcal{S}=\int\mathcal{L}d\tau=\int\left(\dot{x}^{\mu}p_{\mu}-\frac{\lambda}{2}F(\mathbf{p})\right)=\int\left(\dot{x}^{\mu}p_{\mu}-\frac{\lambda}{2}\left(g^{\mu\nu}(\mathbf{p})\,p_{\mu}\,p_{\nu}-m^2\right)\right)
\end{equation}
where the MDR \cref{Finslermdr} has been used.\\
Varying the Lagrangian $\mathcal{L}$ with respect to $\dot{x}^{\mu}$ and the Lagrange multiplier $\lambda$, the equation of motion can be computed:
\begin{align}
&\frac{\partial\mathcal{L}}{\partial\dot{x}^{\mu}}=0\Rightarrow\frac{d}{d\tau}\frac{\partial\mathcal{L}}{\partial\dot{x}^{\mu}}=0\Rightarrow\frac{dp_{\mu}}{d\tau}=0\notag\\
&\frac{\partial\mathcal{L}}{\partial\lambda}=0\Rightarrow g^{\mu\nu}(\mathbf{p})\,p_{\mu}\,p_{\nu}-m^2=0.
\end{align}
The velocity can finally be computed varying the Lagrangian with respect to the momentum $p_{\mu}$ and applying the Euler's theorem to the 0-degree homogeneous metric:
\begin{equation}
\frac{\partial\mathcal{L}}{\partial p_{\mu}}=0\Rightarrow\dot{x}^{\mu}=\lambda\,g^{\mu\nu}(\mathbf{p})\,p_{\nu}+\frac{\lambda}{2}\frac{\partial g^{\alpha\beta}(\mathbf{p})}{\partial p_{\mu}}\,p_{\alpha}\,p_{\beta}=\lambda\,g^{\mu\nu}(\mathbf{p})\,p_{\nu}.
\end{equation}
From this last equation a duality relation connecting velocity and momentum can be determined, posing $\lambda=1/m$:
\begin{equation}
\label{velmom}
\dot{x}^{\mu}=\lambda\,g^{\mu\nu}(\mathbf{p})\,p_{\nu}=\frac{1}{m}\,g^{\mu\nu}(\mathbf{p})\,p_{\nu}\;\Longleftrightarrow\; p_{\mu}=m\,g_{\mu\nu}\,\dot{x}^{\nu}.
\end{equation}
This geometric construction is made in the context of Hamilton-Finsler geometry, starting from the Lagrangian function. In this context, it is possible to find a duality relation between the coordinate space and the momentum space via \cref{velmom}. This result can be compared with the construction made starting from a Hamiltonian function, as in the work \cite{Pfeifer}. Even in this case, the origin of the modified geometry is a 0-degree homogeneous function defined in the momentum space.

\subsection{Non universal scenario in HMSR}
\label{nusHMSR}
The model is constructed considering the possibility of different corrections for each type of interacting particle.
The modified kinematics of the different interacting particle species are correlated employing the vierbein as projectors on a common Minkowski support space. The following graph illustrates how to correlate the geometries related to different probes (local tangent spacetimes), using the vierbein:
\[\begin{tikzcd}
         (TM,\,\eta_{ab},\,p) \arrow{d}{e(p)} \arrow{rr}[swap]{\Lambda} && (TM,\,\eta_{ab},\,q)\arrow{d}[swap]{\overline{e}(q)} \\
        (T_{x}M,\,g_{\mu\nu}(p)) \arrow{rr}[swap]{\widetilde{e}\circ\Lambda\circ e^{-1}} && (T_{x}M,\,\widetilde{g}_{\mu\nu}(q))
\end{tikzcd}\]
The need to perform calculations related to the interactions of different particles implies the introduction of a generalized inner product with the associated metric:
\begin{equation}
\label{c46}
G=\left(
    \begin{array}{cc}
      g^{\mu\nu}(p) & e^{a\mu}(p)\tilde{e}_{a}^{\;\beta}(q) \\
      \widetilde{e}^{a\alpha}(q)e_{a}^{\;\nu}(p) & \widetilde{g}^{\alpha\beta}(q) \\
    \end{array}
  \right)
\end{equation}
where the momenta $p$ and $q$ are associated to different particle species. The generalized Lorentz transformations acquire the form:
\begin{equation}
\label{c47}
\Lambda=
\left(
  \begin{array}{cc}
    \Lambda_{\mu}^{\;\mu'} & 0 \\
    0 & \tilde{\Lambda}_{\alpha}^{\;\alpha'} \\
  \end{array}
\right)
\end{equation}
where the Lorentz transformations related to different spacetimes are coupled. To calculate the physical quantities of interest, an inner product must be defined that associates the momenta of various particles, and this can be done in the following form:
\begin{align}
\label{c48}
&\langle p+q|p+q \rangle=
\left(
  \begin{array}{cc}
    p & q \\
  \end{array}
\right)
\left(
    \begin{array}{cc}
      g^{\mu\nu}(p) & e^{a\mu}(p)\tilde{e}_{a}^{\;\beta}(q) \\
      \tilde{e}^{a\alpha}(q)e_{a}^{\;\nu}(p) & \tilde{g}^{\alpha\beta}(q) \\
    \end{array}
\right)
\left(
  \begin{array}{c}
    p \\
    q \\
  \end{array}
\right)=\notag\\
&\qquad=(p_{\mu}\,e_{a}^{\,\mu}(p)+q_{\mu}\,\tilde{e}_{a}^{\,\mu}(q))\,\eta^{ab}\,(p_{\nu}\,e_{b}^{\,\nu}(p)+q_{\nu}\,\tilde{e}_{b}^{\,\nu}(q)).
\end{align}
In the context of HMSR, the minimal extension of the particle SM envisions the inclusion of various QG perturbations for different interacting particle species. Indeed, most of the predicted phenomenological effects arise from these diverse corrections, exemplified by phenomena such as the GZK (Greisen-Kuzmin-Zatsepin) cut-off phenomenon observed in cosmic rays\cite{Torri1,Torri2,Torri3}.

\section{Standard Model Extension}
The most complete and coherent EFT framework to study the LIV phenomenology is the Standard Model Extension (SME). In this EFT the SM of particle physics is amended supplementing all the possible LIV operators that preserve the usual gauge symmetry $SU(3)\times SU(2)\times U(1)$ \cite{SME,SMECPT,Russell}. The SME formulation is also conceived in order to preserve microcausality, positive energy and four-momentum conservation laws. Moreover, the quantization principles are conserved, such as to guarantee the existence of Dirac and Schr\"odinger equations in the correct energy limit. The modifications introduced in the SME are generated by coupling the standard matter fields with background tensors. These tensors are generated by supposed non-zero vacuum expectation values of some string operators and their constant non-dynamical nature breaks the LI. In this model the covariance is violated for active coordinate transformations, that is the change of coordinates related to the interacting particles.\\
Given the breadth of scenarios offered by the SME for testing covariance, it is imperative to maintain conciseness. Therefore, we focus our investigation solely on the domain of neutrino research.
In the SME formalism, the Lagrangian formulation is approached perturbatively, treating the QG contributions as tiny modifications to the standard physics pattern:
\begin{equation}
\mathcal{L}=\mathcal{L}_{0}+\mathcal{L}_{LIV}.
\end{equation}
The most general Lagrangian for the neutrino sector in the minimal SME has the explicit form:
\begin{equation}
\mathcal{L}=\frac{i}{2}\overline{\psi}_{a}\Gamma^{\mu}_{ab}\partial_{\mu}\psi_{b}-M_{ab}\overline{\psi}_{a}\psi_{b}+h.c.
\end{equation}
where the $\Gamma$ and $M$ matrices are constructed using all the perturbative terms introduced in SME:
\begin{align}
\label{CPTmass}
&\Gamma^{\mu}_{ab}=\gamma^{\mu}_{ab}+c^{\mu\nu}_{ab}\gamma_{\nu}+d^{\mu\nu}_{ab}\gamma_{\mu}\gamma_{5}+e^{\mu}_{ab}+i\,f^{\mu}_{ab}\gamma_{5}+\frac{1}{2}g^{\alpha\beta\mu}_{ab}\sigma_{\alpha\beta}\notag\\
&M_{ab}=m_{ab}+i\,m_{5ab}\gamma_{5}+a^{\mu}_{ab}\gamma_{\mu}+b^{\mu}_{ab}\gamma_{5}\gamma_{\mu}+\frac{1}{2}H^{\mu\nu}_{ab}\sigma_{\mu\nu}
\end{align}
where the coefficients are CPT-even or CPT-odd depending on the parity of the number of Lorentz coefficients.\\
Since the LIV CPT-even and CPT-odd corrections can be studied considering a subgroup of the SME parameters, in this work we will consider only the contributions given by the LIV Lagrangian:
\begin{equation}
\mathcal{L}_{LIV}=-a_{\mu}\overline{\psi}_{L}\gamma^{\mu}\psi_{L}-c_{\mu\nu}\overline\psi_{L}\gamma^{\mu}\partial^{\nu}\psi_{L}.
\end{equation}
The first term proportional to the background field $a_{\mu}$ violates the CPT symmetry and the LI as a consequence. The second term proportional to $c_{\mu\nu}$ violates the Lorentz covariance, but preserves the CPT symmetry.
In order to describe the propagation of the neutrino, especially the phenomenon of oscillations, it is useful to switch to the Hamiltonian approach, writing the effective Hamiltonian function as:
\begin{equation}
\mathcal{H}_{ab}=E\delta_{ab}+\frac{1}{2E}M^2_{ab}+\frac{2}{E}\left([a^{\mu}p_{\mu}]_{ab}-[c^{\mu\nu}p_{\mu}p_{\nu}]_{ab}\right).
\end{equation}
In the high energy limit the effective Hamiltonian can be written as:
\begin{equation}
\label{effneuham}
\mathcal{H}_{eff}=\mathcal{H}_{0}+\mathcal{H}_{LIV}^{CPT-odd}+\mathcal{H}_{LIV}^{CPT-even}=\mathcal{H}_{0}+
\left[
\left(
  \begin{array}{ccc}
    a_{ee} & a_{e\mu} & a_{e\tau} \\
    a^{*}_{e\mu} & a_{\mu\mu} & a_{\mu\tau} \\
    a^{*}_{e\tau} & a^{*}_{\mu\tau} & a_{\tau\tau} \\
  \end{array}
\right)-
\frac{3}{4}E
\left(
  \begin{array}{ccc}
    c_{ee} & c_{e\mu} & c_{e\tau} \\
    c^{*}_{e\mu} & c_{\mu\mu} & c_{\mu\tau} \\
    c^{*}_{e\tau} & c^{*}_{\mu\tau} & c_{\tau\tau} \\
  \end{array}
\right)
\right].
\end{equation}
In the case of SME the Lorentz covariance is violated, hence, there is no necessity for the establishment of mechanisms designed to integrate corrections pertaining to distinct geometrical structures.

\section{GRB and Supernova neutrinos}
\label{GRB}
\begin{figure}[ht]
\begin{centering}
\includegraphics[scale=0.38]{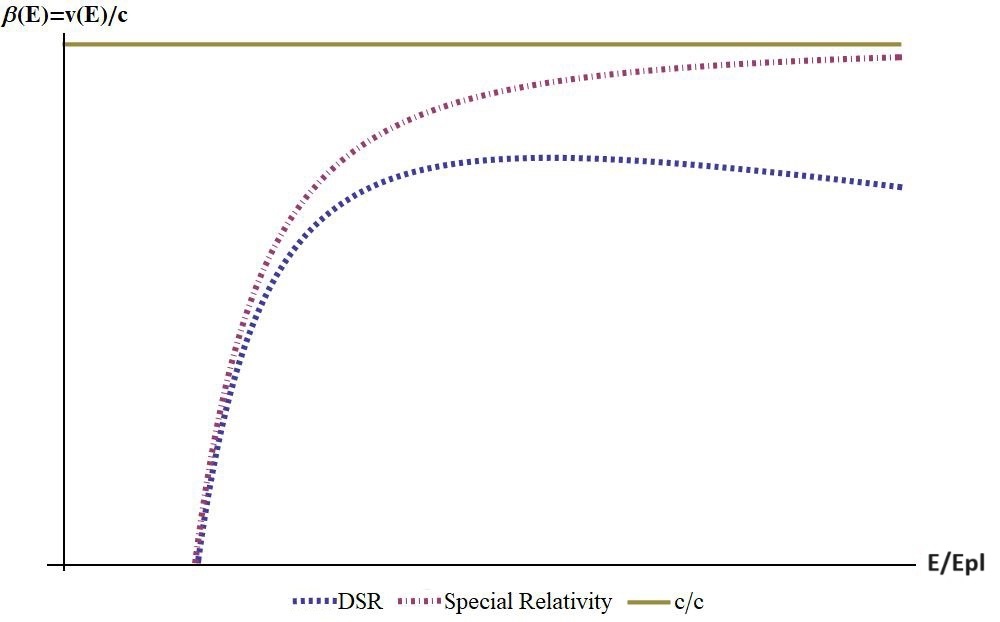}
\caption{Comparison of the Lorentz factor $\beta=v/c$ as a function of the energy with the factor $\beta=c/c$ (green continuous line): DSR (blue dashed line) vs Special Relativity (red dash dotted line).}
\label{fig1}
\end{centering}
\end{figure}

One of the main channels for the detection of QG signatures involves measuring the time of flight of astroparticle neutrinos with different energies in a multi-messenger approach. Indeed, QG perturbations can affect the in-vacuum dispersion relations introducing an energy dependence of the particle velocities. In the context of DSR theories the foreseen energy dependence of the particle velocity can be obtained assuming a positive QG correction in \cref{mdrdef}. Starting from the MDR the energy can be obtained:
\begin{equation}
E\simeq\sqrt{p^2\left(1+\delta\cdot p/M_{P}\right)+m^2}
\end{equation}
and applying the Hamilton's equation the velocity is computed:
\begin{equation}
v(E)=\frac{\partial}{\partial p}E\simeq\frac{p+3/2\cdot\delta\cdot p^2/M_{P}}{\sqrt{p^2+\delta\cdot p^{3}/M_{P}+m^2}}
\end{equation}
where $\delta$ represents the constant encoding the magnitude of the QG perturbations for this phenomenon.\\
From \cref{fig1}, it is possible to deduce that the velocity decreases as energy increases, demonstrating that the flight time of more energetic particles can be greater. The plot is obtained with a suitable choice of parameters (i.e. the particle mass $m$ and the Q.G. perturbation parameter $\delta$) aimed at amplifying the effect. The quantities indicated in the plot are obtained in dimensionless units: the plot reports the Lorentz factor $\beta=v(E)/c$ (appearing in ordinates) as a function of the energy (in abscissa), measured in units of the Planck energy $E_{Pl}=\sqrt{\dfrac{\hbar c^{5}}{G}}$.
The positive corrections in \cref{mdrdef} are not taken into consideration, as they would result in super-luminal neutrinos. This hypothesis is ruled out since such particles would undergo pair production processes, dissipating energy during propagation \cite{Cohen,Vissani}. Moreover, the hypothesis of super-luminal neutrinos would contradict known physics, particularly in scenarios such as pion decay \cite{Vissani2}.\\
In this study, we will consider the MDR \cref{mdrdef} in order to analyze the time arrival of different energies astroparticles \cite{Amelino1,Amelino2,Ellis,Jacob,Wei}. Thanks to the peculiar form of the dispersion relation, it is simple to calculate the QG contribution to the flight time as a function of the particle energy:
\begin{equation}
\label{dt}
\Delta t=\delta\,D(z)\,\frac{E}{M_{Pl}}
\end{equation}
where $\delta$ is the constant introduced before and $D(z)$ is a function of the redshift $z$ related to the astrophysical source of the detected particle:
\begin{equation}
D(z)=\int_{0}^{z}d\zeta\frac{1+\zeta}{H_{0}\sqrt{\Omega_{\Lambda}+\Omega_{m}(1+\zeta)^3}}
\end{equation}
here, as usual, $H_{0}$, $\Omega_{\Lambda}$ and $\Omega_{m}$ denote respectively the Hubble constant, the cosmological parameter and the matter fraction \cite{Planck}. The difference in time travel represents the delay accumulated by particles of the same species as a function of energy compared to what is predicted in the classical propagation scenario. This time delay, as described in \cref{dt}, exhibits a linear dependence on the particle energy. Obviously, the linear dependence is still valid for other constant $\delta$ values. In the following table we report the computation of some time delay values as a function of the neutrino energy \cref{dt}, computed for the redshift value $z=1$ and QG perturbation constant $\delta=1$:
\begin{table}[H]
\center
\begin{tabular}{|c|c|l}
  \hline
  $E_{\nu}$ (TeV) & $\Delta t$ (s)\\ \hline \hline
  $10$  & $0.5\times 10^{3}$ \\ \hline
  $20$  & $10^{3}$ \\ \hline
  $100$  & $5\times 10^{3}$  \\ \hline
  $200$  & $10^{4}$  \\ \hline
  \hline
\end{tabular}
\end{table}
Detecting such an effect in the astroparticle sector can be straightforward, thanks to the high energies and propagation paths involved. Gamma-ray bursts (GRB) are a suitable class of phenomena for this study, as they emit nearly simultaneous photons with varying energies and likely accelerate neutrinos \cite{Waxman,Biehl,Rachen,Guetta,Ahlers} to the extremely high energies required for this investigation \cite{Amelino1,Amelino2,Ellis,Jacob,Wei}. The effect can be detected looking at the time delay accumulated by the most energetic particles.\\
The emission spectrum of neutrinos accelerated in a GRB is described by a double broken power law \cite{IceCube2014}:
\begin{equation}
\phi_{\nu}(E_{\nu})=\phi_{0}\cdot
\begin{cases}
\begin{aligned}
& \epsilon_{b}^{-1}E_{\nu}^{-1} &E_{\nu}\leq\epsilon_{b}\\
& E_{\nu}^{-2} &\epsilon_{b}\leq E_{\nu}\leq10\cdot\epsilon{b}\\
& (10\,\epsilon_{b})^{2}E_{\nu}^{-4} &E_{\nu}\geq10\cdot\epsilon_{b}
\end{aligned}
\end{cases}
\end{equation}
where $\phi_{0}$ is the spectral normalization flux and $\epsilon_{b}$ is the break energy. Some properties of the GRB prompt phase of the neutrino emission are not well known and still under debate, such as the time dependence of the emission spectrum \cite{Kimura,Kimura2}. Despite this dependence on the model considered for neutrino emission phase, some research works foresee the possibility to conduct this kind of investigation, finding a good agreement with the expected linear dependence of the time delay on the neutrino energy \cite{Amelino1,Amelino2,Ellis,Jacob,Wei,Guetta,Huang,Zhang}.\\
Another candidate sector where this research can be conducted is the detection of neutrinos generated during a star collapse and the following Supernova (SN) explosion \cite{Mirizzi,Janka}. During the stellar collapse preceding the SN explosion, neutrinos are emitted in three different stages. Initially, the neutronization phenomenon occurs, generating electron neutrinos and neutrons through the fusion of protons and electrons. The electron neutrinos created in this stage become trapped behind the shock wave of collapsing stellar matter and are released only when the matter density decreases sufficiently. In the subsequent stage, known as accretion, neutrinos of all flavors are produced. In the third stage the newly formed neutron star cools, reducing its binding energy and emitting all flavor neutrinos. The different neutrino flavors exhibit distinct differential fluxes, which can be expressed as functions of time and energy:
\begin{equation}
\frac{d^2\Phi_{\alpha}}{dt\,dE}=\frac{L_{\alpha}(t)}{4\,\pi\,d^2}\frac{f_{\alpha}(t,\,E)}{\langle E_{\alpha}(t)\rangle}
\end{equation}
where $\Phi_{\alpha}$ represents the original neutrino flux, $L_{\alpha}(t)$ is the luminosity, $d$ is the SN-detector distance and $f_{\alpha}(t,\,E)$ is the energy spectrum for each $\alpha$ neutrino flavor.\\
Each neutrino flavor spectrum can be parameterized using the function:
\begin{equation}
\label{spectrum}
f_{\alpha}(t,\,E)=\lambda_{\alpha}\left(\frac{E}{\langle E_{\alpha}(t)\rangle}\right)^{\beta_{\alpha}}\exp{\left(-\frac{(\beta_{\alpha}(t)+1)\,E}{\langle E_{\alpha}(t)\rangle}\right)}
\end{equation}
here $\beta_{\alpha}$ represents the model-dependent pinching parameter, accounting for the thermal neutrino spectrum, and
$\langle E_{\alpha}(t)\rangle$ is the mean neutrino energy.\\
As discussed above, neutrinos with varying energies can accumulate an energy-dependent time delay, as described by \cref{dt}. This phenomenon can be detected analyzing the energy spectrum, for instance by dividing it, as shown in \cref{spectrum}, into different bins and evaluating the accumulated delays of the neutrinos within each selected energy interval.\\
All the results reported in this section are phenomenological consequences of the QG induced modifications of the free particle dispersion relation. It is interesting to highlight that similar results can be obtained in effective scenarios derived from string theory \cite{Ma,Ma2} and LQG \cite{Ma3}.

\section{Oscillations in atmospheric neutrinos}
\begin{figure}[ht]
\begin{centering}
\includegraphics[scale=0.52]{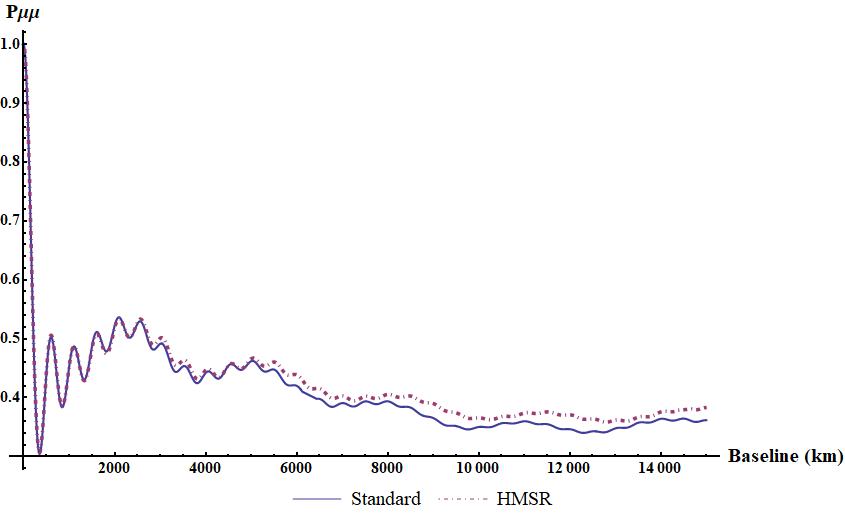}
\caption{Comparison of the integrated atmospheric neutrino survival probability (energy range: 500 Mev  -3 GeV) as a function of the baseline: standard case (blue continuous line) vs HMSR (red dash dotted line).}
\label{atmosphericHMSR}
\end{centering}
\end{figure}

\begin{figure}[ht]
\begin{centering}
\includegraphics[scale=0.52]{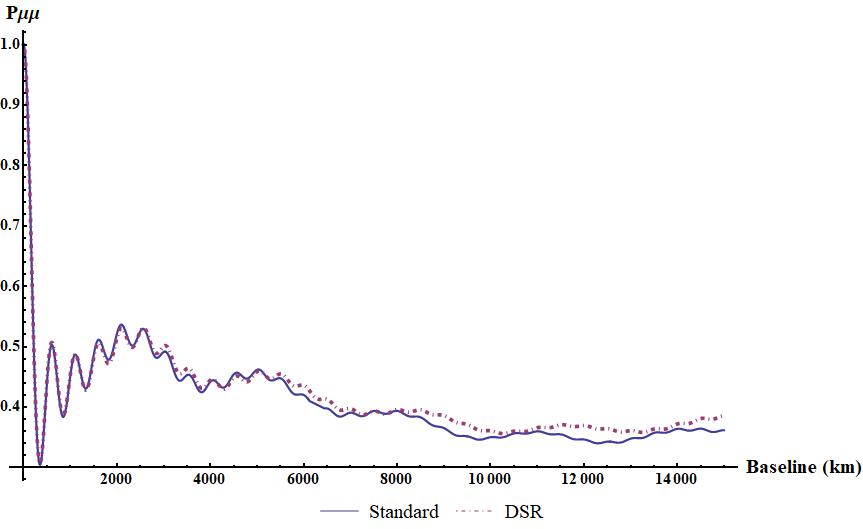}
\caption{Comparison of the integrated atmospheric neutrino survival probability (energy range: 500 Mev  -3 GeV) as a function of the baseline: standard case (blue continuous line) vs DSR (red dash dotted line).}
\label{atmosphericDSR}
\end{centering}
\end{figure}

Neutrinos can also serve as probes for the potential non-universality of gravitational interaction resulting from hypothesized QG perturbations. Indeed, the oscillation phenomenon can be influenced by the introduction of dispersion relation modifications that depend on the particle species. The QG-induced modification of oscillations has been extensively investigated in the context of both atmospheric and astrophysics neutrinos by the SME \cite{IceCubeLIV,IceCubeLIV2,Mewes,Diaz,Torri4}. It is possible to determine Planck-scale perturbations in neutrino oscillations also within the framework of the Generalized Uncertainty Principle (GUP) formalism \cite{Sprenger,Sprenger2}, by introducing a generalized commutator of the type:
\begin{equation}
[x^{\mu},\,p_{\nu}]=i(1+f(|p|^2))\delta^{\mu}_{\,\nu}.
\end{equation}
This GUP commutator can be compared with the modified commutation rules obtained in the context of the HMSR model \cref{commutationsHMSR}.\\
In this work we introduce the possibility to investigate such a phenomenon in the context of modifications of the covariance, considering DSR and HMSR models. Thanks to the results of \cref{nusDSR,nusHMSR}, it is possible to contemplate non-universal scenarios for the QG motivated extensions of the SM of particle physics for Lorentz modifying models. In the case of particle-depending MDR, the various neutrino mass eigenstates can be perturbed in distinct ways.\\
Following the HMSR model \cite{Torri4,Torri5} and considering the MDR \cref{Finslermdr}, in the ultra-relativistic particle limit, where $|\vec{p}|\simeq E$, the following relation can be obtained:
\begin{equation}
\label{approx1}
|\vec{p}|\simeq\left(1-f_{j}\left(\frac{|\vec{p}|}{E}\right)\right)E+\frac{m_{j}^2}{2E}\simeq \left(1-\epsilon_{j}\right)E+\frac{m_{j}^2}{2E}
\end{equation}
for each mass eigenstate $j$. The constant $\epsilon_{j}$ defines the first order of the perturbative series expansion of the function $f_{j}\left(\frac{|\vec{p}|}{E}\right)=\epsilon_{j}\cdot\left(\frac{|\vec{p}|}{E}\right)$.\\
Using the previous equality, the plane wave phase related to each neutrino mass eigenstate can be computed:
\begin{equation}
\phi_{j}=E\,t-E\,L+\epsilon_{j}\,E\,L-\frac{m_{j}^{\,2}}{2\,E}=\left(\epsilon_{j}E-\frac{m_{j}^{\,2}}{2\,E}\right)L
\end{equation}
utilizing the natural measure units for which $t\,=\,L$. Hence, the phase difference of two mass eigestates can be written in the form:
\begin{equation}
\label{phase}
\Delta\phi_{jk}=\left(\frac{\Delta m_{jk}^2}{2\,E}-\epsilon_{jk}\,E\right)\,L
\end{equation}
here $\Delta m_{jk}^2$ is the usual mass squared difference and $\epsilon_{jk}$ represents the disparity of the correction constants: $\epsilon_{jk}=\epsilon_{j}-\epsilon_{k}$. Therefore, allowing diverse corrections for each particle, it is possible to foresee detectable effects caused by the QG perturbations. As a result the phase is modified by the introduction of a term proportional to the particle energy $E$ multiplied by the propagation length.\\
An analogous computation can be conducted in the context of DSR theories. Starting from the dispersion relation \cref{mdrdef} one can obtain the following:
\begin{equation}
|\vec{p}|\simeq\left(1-\delta\frac{E}{M_{P}}\right)E+\frac{m_{j}^2}{2E}
\end{equation}
where $\delta$ is a proportional constant which encodes the geometrical structure of QG perturbations. In this case the phase related to each neutrino mass eigenstate can be written as:
\begin{equation}
\phi_{j}=E\,t-E\,L+\delta_{j}\,\frac{E^{2}}{M_{P}}\,L-\frac{m_{j}^{\,2}}{2\,E}\simeq\left(\frac{\delta_{j}}{M_{P}}\,E^{2}-\frac{m_{j}^{\,2}}{2E}\right)\,L
\end{equation}
where $\delta_{j}/M_{P}$ is the constant governing the magnitude of the QG corrections. The QG perturbation depends once again on the particle species, specifically on the distinct mass eigenstate. Consequently, we can calculate the difference between the phases of two mass eigenstates, mirroring the outcome obtained in \cref{phase}:
\begin{equation}
\Delta\phi_{jk}=\left(\frac{\Delta m_{jk}^2}{2\,E}-\frac{\delta_{jk}}{M_{P}}\,E^{2}\right)\,L
\end{equation}
In this context, $\delta_{jk}$ represents the disparity between the correction factors of the two mass eigenstates under consideration, with $\delta_{jk}=\delta_{j}-\delta_{k}$. In contrast to the preceding scenario, in this case, the QG correction exhibits a proportionality to $E^2$.\\
The QG induced modifications of the phase differences can influence the final form of the oscillation probability \cref{prob} impacting the oscillation pattern.\\
It is important to emphasize that the corrections derived here, within the framework of DSR and HMSR models, are CPT-even. These corrections are motivated by the hypothesis that each mass eigenstate interacts differently with the background due to the postulated QG perturbations. Another crucial point to highlight is that the introduced QG corrections cannot explain the neutrino oscillation phenomenon without the inclusion of masses. This is because the QG perturbation term is proportional to the particle energy or the squared energy, differing from the typical functional form of the oscillation phase. Consequently, the QG correction must be only a perturbation to the usual phase ruling oscillations.\\
This effect can be detected in the atmospheric sector by comparing the expected fluxes of different-flavor neutrinos.
In \cref{atmosphericHMSR}, we juxtapose the integrated probabilities between the standard scenario and the QG modified one, as derived within HMSR framework. Specifically, we calculate the integrated probability of muonic atmospheric neutrinos within the energy range of 500 MeV to 3 GeV. To obtain the muonic survival probability, we integrate the predicted analytical spectrum, taken from \cite{atmospheric} multiplied by the survival probability, a function of both energy and baseline. We have assumed that the QG dimensionless correction factors are equal for every pair of mass eigenstates selected, meaning $\epsilon_{jk}=10^{-23}$ for every choice of eigenstates $j$ and $k$ \cref{phase}. This value has been intentionally chosen to be sufficiently large in order to accentuate the visibility of the effect. The probability is plotted as a function of the baseline, which spans the range relevant to atmospheric neutrinos.
In \cref{atmosphericDSR}, we present a plot similar to the previous one, but generated within the framework of DSR theory. In this plot, we report the integrated probability, maintaining consistent QG disparity factors for every pair of mass eigenstates, as before, i.e., $\alpha_{jk}=5\times10^{-33}$ for every $j$ and $k$ eigenstate. In this case, a smaller value for the QG correction factor was chosen because the QG correction terms are proportional to $E^{2}$.\\
Future experiments will offer unprecedented precision in reconstructing arrival direction, energy, and flavor of atmospheric neutrinos, enabling the comparison of collected data with predictions. The detection of an excess of one flavor caused by a modification of the survival probability may therefore demonstrate the non-universality of gravitational interaction.

\section{CPT violation}
The CPT symmetry is deeply embedded in the SM of particle physics and it is closely linked to the Lorentz covariance and the unitarity of the formulation of the relativistic QFT. Since breaking this symmetry implies LIV \cite{Greenberg}, it is plausible to attribute this supposed violation to QG induced decoherence effects. In this case, CPT violation is expected to be suppressed at the Planck scale, with its magnitude naively estimated as $\langle v^2\rangle/M_{P}\simeq 10^{-5}$ eV$^{-1}$ \cite{Mavromatos,Torri6}.\\
In a CPT violating scenario the QG-induced corrections of neutrinos and antineutrinos can be different. Particles and antiparticles may have different propagation properties. Starting from the MDR it is possible to obtain the different group velocity, predicting the possibility of super-luminal particles or antiparticles. This possibility is already be ruled out in the \cref{GRB}, since superluminal neutrinos may violate the known pion decay physics \cite{Vissani2}.\\
Ruled out the possibility of super-luminal neutrinos, CPT symmetry can imply that an antiparticle can be viewed as a particle with the same mass propagating backward in time. One way to break CPT symmetry is by introducing different masses for particles and antiparticles, as foreseen in the CPT-odd extension of the SME \cite{SMECPT}, this is because the mass operator \cref{CPTmass} includes CPT-odd terms. Admitting different masses for particles and antiparticles implies a violation of the theory unitarity.
Within the framework of the SME, which accommodates scenarios with CPT violation, it is conceivable that neutrinos and antineutrinos exhibit differences in their masses and oscillation parameters. An effective approach to detect potential CPT violations in this sector involves a comparative analysis of various experiments, aiming to identify discrepancies in the oscillation parameters between neutrinos and antineutrinos. Currently, the best constraints on the differences in oscillation parameters between neutrinos and antineutrinos are obtained by comparing data from accelerator neutrino experiments with reactor antineutrino experiments \cite{Barenboim}:
\begin{align}
&|\Delta m_{21}^2-\Delta\overline{m}_{21}^2|<3.7\times10^{-5}\;\text{eV}^2\notag\\
&|\Delta m_{31}^2-\Delta\overline{m}_{31}^2|<2.4\times10^{-4}\;\text{eV}^2\notag\\
&|\sin^2{\theta_{12}}-\sin^2{\overline{\theta}_{12}}|<0.187\notag\\
&|\sin^2{\theta_{13}}-\sin^2{\overline{\theta}_{13}}|<0.029\notag\\
&|\sin^2{\theta_{23}}-\sin^2{\overline{\theta}_{23}}|<0.19.
\end{align}
To enhance this synergic analysis, a future perspective involves the data from solar neutrinos, thereby capitalizing information from both accelerator and solar neutrino fluxes, in contrast to reactor antineutrinos.
This endeavor can be realized through the next generation neutrino experiments under construction, such as the Jiangmen Underground Neutrino Observatory (JUNO), Hyper-Kamiokande and the Deep Underground Neutrino Experiment (DUNE).\\
JUNO \cite{JUNO1,JUNO2} represents a medium baseline multipurpose reactor experiment, designated to investigate various aspects of neutrino physics, such as the discrimination of neutrino mass ordering and the precise measurement of the solar neutrino and the reactor antineutrino oscillation parameters. The experiment will employ a central detector filled with 20,000 tons of liquid scintillator, surrounded by 15,000 new-generation photomultipliers. This setup will offer an unprecedented 3\% energy resolution for a neutrino beam at 1 MeV, providing the requisite sensitivity for achieving JUNO's primary objectives. The main detector will be situated approximately 52 km away from eight reactors at two different nuclear power plants, boasting a combined output of approximately $7.5$ GW.\\
Hyper-Kamiokande \cite{HYPERK} represents the next generation in water Cherenkow detectors, with a fiducial volume $8$ times larger than its predecessor Super-Kamiokande. This experiment will investigate not only the solar and atmospheric sector but also long-baseline accelerator neutrinos.\\
DUNE \cite{DUNE,DUNE2,DUNE3}, on the other hand, takes another approach as a next generation neutrino observatory, using the charged interaction of electronic neutrinos with liquid Argon: $\nu_{e}+^{40}\text{Ar}\rightarrow e^{-}+^{40}K$. This experiment will be sensitive to the detection of solar neutrinos with energies exceeding $9$ MeV.\\
An experimental challenge may arise from potentially confusing contributions caused by hypothesized CPT-odd violations with those caused by the supposed NSI, briefly discussed in the next section.

\section{Non Standard Interactions}
Since the discovery of the oscillation phenomenon and the associated masses, which were not predicted by the Standard Model, neutrino physics has been a rich field for investigating new physics. Indeed, some new physical models offer phenomenological predictions that can be tested in the neutrino sector, as they allow for the possibility of NSI. For example, NSI are predicted by certain super-symmetric theories or models conceived to explain the hypothesized presence of dark matter. The induced modifications are motivated by amending the usual SM symmetries, but are not strictly caused by QG perturbations. These exotic models are named vectorial NSI since they modify the charged current interactions of neutrinos with electrons, affecting the neutrino matter interactions.\\
Following an EFT approach to the NSI, the neutrino interaction lagrangian density can be written as:
\begin{equation}
\mathcal{L}_{NSI}=-2\sqrt{2}G_{F}\sum_{\alpha\,\beta}\,\epsilon_{\alpha\beta}^{ll'\,X}\left(\overline{\nu}_{\alpha}\gamma_{\mu}P_{L}\nu_{\beta}\right)\left(\overline{l}\gamma^{\mu}P_{X}l'\right)
\end{equation}
where the indices $\alpha,\,\beta$ represent the incoming and outgoing neutrino flavors, $l$ and $l'$ are the incoming and outgoing interacting leptons with the same charge, but not necessarily the same flavor, and $X$ specifies the $ll'$ interaction lepton chirality. The hermitian NSI matrix $\epsilon_{\alpha\beta}^{ll'\,X}$ enforces the meaning of NSI, allowing for flavor non-diagonal or left-right chirality interactions. Using the hamiltonian picture, the Hamilton operator associated to NSI interactions is written in the form:
\begin{equation}
\label{nsi}
\mathcal{H}_{NSI}=2\sqrt{2}\,G_{F}
\left(
  \begin{array}{ccc}
    \epsilon_{ee} & \epsilon_{e\mu} & \epsilon_{e\tau} \\
    \epsilon_{\mu e} & \epsilon_{\mu\mu} & \epsilon_{\mu\tau} \\
    \epsilon_{\tau e} & \epsilon_{\tau\mu} & \epsilon_{\tau\tau} \\
  \end{array}
\right)
\end{equation}
where the entries are subject to the hermiticity constraint $\epsilon_{\alpha\beta}=\epsilon_{\beta\alpha}^{*}$.\\
NSI affects the neutrino interaction with matter and can therefore influence the experimental results modifying the neutrino detector behaviour. In this context, it's essential to highlight the similarity between the Hamiltonian operator of NSI \cref{nsi} and the Lorentz Invariance Violation (LIV) CPT-odd operator \cref{effneuham} as envisaged in the SME framework. Due to the similar mathematical formulation of these two operators, the perturbations introduced by NSI can mimic the effects of QG-induced CPT-odd effects. We highlight that the SME includes all terms that violate Lorentz symmetry and eventually the CPT symmetry, even those not caused by QG effects. In the SME formulation, for instance, the usual gauge symmetry of the particle SM is preserved. What sets apart NSI is that this type of interaction, unlike the perturbations introduced in the SME, have a different nature since they can violate the particle SM, introducing exotic interactions not caused by QG effects. Therefore, disentangling these contributions is of fundamental importance in every experimental context. This task is really complicated and effectively separating the two contributions within the framework of a single experiment appears to be an exceptional challenge \cite{Barenboim}. This objective can be achieved through complementary studies involving data from different detectors. For example, comparing the results of day/night asymmetry in neutrino matter interactions \cite{JUNO3} and those obtained in the context of Non-Standard Interactions (NSI) in the solar sector \cite{BOREXINO} with the constraints imposed on LIV \cite{Russell}.

\section{Quantum decoherence}
The CPT symmetry breaking caused by gravitational effects can induce violations of the unitarity of the theory \cite{Torri6}, which implies decoherence effects in the propagation of neutrino wave packets. Indeed, gravitational effects can affect the predictability of the future. For instance, black holes and their event horizons can limit the physics knowledge of every observer \cite{Mavromatos}. In the context of the space-time foam idea the evolution of pure states can undergo decoherence effects caused by the non commutativity of coordinates or the fluctuations of the metric tensor \cite{ArzanoDec,DEsposito}. In this section we briefly review the possible implications of QG-induced decoherence in neutrino oscillation sector. The introduction of QG space-time perturbations can affect the standard oscillation pattern introducing a damping exponential factor, which quenches the oscillation phenomenon without altering the neutrino total flux.\\
The framework adopted to describe the quantum decoherence effects is the wave packet formalism. In this context the neutrino evolution can be described using the density matrix of different states $\rho(t)$, and the oscillation probability results:
\begin{equation}
P(\nu_{\alpha}\rightarrow\nu_{\beta},\,t)=Tr\bigg[\rho(t)|\nu_{\alpha}\rangle\langle\nu_{\alpha}|\bigg]
\end{equation}
with the time evolution of the density matrix written using the Hamiltonian:
\begin{equation}
\rho(t)=e^{-iHt}|\nu_{\beta}\rangle\langle\nu_{\beta}|=\sum_{i,\,j}U^{*}_{\beta\,i}U_{\beta\,j}\int\,d^{3}p\,d^{3}q\,\psi_{i}(\mathbf{p})\,\psi^{*}_{j}(\mathbf{q})e^{-i(E_{i}(\mathbf{p})-E_{j}(\mathbf{q}))t}|\nu_{i}\rangle\langle\nu_{j}|
\end{equation}
where the different neutrino states are written in the form $|\nu_{i}(\mathbf{p})\rangle=\psi_{i}(\mathbf{p})|\nu_{i}\rangle$ and the PMNS matrix is used.\\
By collecting all the previous results, and using the relation:
\begin{equation}
|\nu_{\alpha}(\mathbf{q})\rangle\langle\nu_{\alpha}(\mathbf{q})|=\sum_{i,\,j}U^{*}_{\alpha\,i}U_{\alpha\,j}\phi^{*}_{i}(\mathbf{q})\phi_{j}(\mathbf{q})|\nu_{\alpha}\rangle\langle\nu_{\alpha}|
\end{equation}
we can express the oscillation probability as follows:
\begin{equation}
P(\nu_{\alpha}\rightarrow\nu_{\beta},\,t)=\sum_{i,\,j}U^{*}_{\alpha\,j}U_{\alpha\,i}U^{*}_{\beta\,i}U_{\beta\,j}\int\,d^{3}p\,d^{3}q\,\phi^{*}_{i}(\mathbf{p})\phi_{j}(\mathbf{q})\psi^{*}_{j}(\mathbf{q})\phi_{i}(\mathbf{p})e^{-i(E_{i}(\mathbf{p})-E_{j}(\mathbf{q}))t}
\end{equation}
The decoherence effect can be described using the Lindebland equation formalism \cite{DEsposito,Hellmann}:
\begin{equation}
\frac{\partial\,\rho(t)}{\partial\,t}=\mathfrak{L}[\rho(t)]=-i[H,\,\rho]+\sum_{j}\left(\mathfrak{L}_{j}\,\rho\,\mathfrak{L}_{j}^{\dagger}-\frac{1}{2}\big\{\mathfrak{L}_{j}^{\dagger}\mathfrak{L}_{j},\,\rho\big\}\right).
\end{equation}
where the time evolution operator $\mathfrak{L}$ has been introduced, acting on the density matrix $\rho$ instead of the usual Hamiltonian $H$. In this context the time evolution of the density matrix can be written as:
\begin{equation}
\rho(t)=e^{\mathfrak{L}t}\bigg[|\nu_{\beta}\langle\nu_{\beta}|\bigg]=\sum_{i\,j}U^{*}_{\beta\,i}U_{\beta\,j}\int\,d^{3}p\,d^{3}q\,\psi_{i}(\mathbf{p})\psi^{*}_{j}(\mathbf{q})e^{-i(E_{i}(\mathbf{p})-E_{j}(\mathbf{q}))t}e^{-\mathfrak{L}_{ij}(\mathbf{p},\,\mathbf{q})t}|\nu_{i}\rangle\langle\nu_{j}|
\end{equation}
$\mathfrak{L}_{ij}$ is the model-dependent function encoding the perturbation induced by the Lindebland operator.\\
Under the assumption of neutrino momenta peaked around the average value, the energy can be approximated with the series expansion $E_{i}(p)=E_{i}(\overline{p})+(p-\overline{p})v_{g(i)}$ written as a function of the group velocity $v_{g(i)}=\overline{p}_{i}/E_{i}$. The oscillation probability is amended by the introduction of the Lindebland time evolution, and in every decoherence scenario, it can be expressed as:
\begin{equation}
P(\nu_{\alpha}\rightarrow\nu_{\beta},\,t)=\sum_{i,\,j}U^{*}_{\beta,\,i}U_{\beta\,j}U^{*}_{\alpha\,j}U_{\alpha\,i}e^{\phi_{ij}}e^{-D_{ij}(\mathbf{p},\,\mathbf{q})}
\end{equation}
where $D_{ij}(\mathbf{p},\,\mathbf{q})=\mathfrak{L}_{ij}(\mathbf{p},\,\mathbf{q})/v_{g(ij)}$ is the damping factor introduced by the decoherence effect and $v_{g(ij)}=(v_{g(i)}+v_{g(j)})/2$ is the mean group velocity.\\
The decoherence damping factor can be computed in the context of some QG scenarios, such as:
\begin{itemize}
  \item symmetry deformations (DSR theories): $D_{ij}=\frac{L\,(\Delta m_{ij}^{2})^2}{8v_{g}M_{P}p^{2}}$ \cite{ArzanoDec}
  \item metric perturbations: $D_{ij}=\frac{L\,E^{6}(\Delta m_{ij}^2)^{2}}{4v_{g}M_{P}m_{i}^{4}m_{j}^{4}}$ \cite{Breuer}
  \item fluctuating minimal length: $D_{ij}=\frac{16\,L\,E^{4}(\Delta m_{ij}^2)^2}{v_{g}M_{P}^{5}}$ \cite{Petruzziello}.
\end{itemize}
In all these cases, the damping factor linearly depends on the propagation length of the neutrino beam. As it is illustrated in \cite{DEsposito}, only the metric perturbations can significantly affect the oscillation pattern enough to allow for the effects detection under the condition $D_{ij}\geq1$. The ideal setup to test these predictions is provided by long and medium baseline reactor neutrino experiments \cite{DEsposito}. In these cases, the neutrino flux is not affected, werehas the oscillation probability is modified by the supposed QG caused decoherence. Nevertheless, in this sector, the decoherence effects caused by the different propagation velocities of the different mass eigenstates can interfere with the QG-induced decoherence effects \cite{DEsposito}.\\
Another interesting scenario is the case of astrophysics neutrinos \cite{Stuttard,IceCubeDec}. In this scenario, the loss of coherence caused by QG can manifest determining a loss of unitarity. Indeed, if neutrinos interact with a fluctuating background, different initial quantum wave states can lead to the same final states, potentially violating unitarity. In such a scenario, Planck scale effects are expected to manifest strongly in high-energy particles propagating over long distances.

\section{Conclusion}
In this work, we have explored the possibility of using neutrinos as probes to investigate the supposed quantum structure of spacetime. We presented three distinct research frameworks that enable the study of QG-induced perturbations in various aspects of neutrino physics. These models exhibit a fundamental distinction: the DSR and HMSR theories preserve covariance, albeit in an amended formulation, while the SME explicitly breaks LI.\\
We introduced a mathematical formulation within the context of DSR theories to account for the possibility of non-universal QG corrections, allowing us to search for violations of gravitational interaction. This formulation employs Hopf-algebra and coproduct formalism, establishing a mechanism for connecting different symmetry algebras related to various interacting particles. The HMSR model is conceived from its foundation to consider the potential for particle-dependent QG perturbations.\\
From a phenomenological perspective, we explored various sectors of neutrino physics where hypothesized QG signatures may manifest. Initially, we considered the possibility of detecting neutrinos associated with GRB or SN explosions. Due to their long propagation paths and the wide range of energies spanned by these astroparticles, they offer an ideal sector for searching for QG perturbations. In fact, by assuming MDR that depend on particle energy, we demonstrated that QG could introduce differences in the arrival times of neutrinos with different energies.\\
Next, we investigated the possibility of detecting signatures potentially caused by non-universal corrections in the oscillation probability. In this context, violations of universality may introduce particle-dependent perturbations that may affect the neutrino oscillation pattern. The atmospheric sector is ideal for searching for such effects by looking for deviations from the standard predictions in the survival probability of different neutrino flavors. We have demonstrated the possibility of predicting this effect in both DSR and HMSR research framework.\\
We explored the possibility of detecting CPT violations in the neutrino sector, which can only be examined within the framework of the SME. Phenomenologically, this violation can be seen as an introduced disparity between the masses of particles and their corresponding antiparticles. The particle mass can acquire operator perturbation terms that differ from those related to the corresponding antiparticle. To search for such effects, the ideal setting is once again related to neutrino oscillations, requiring a comparison of neutrino and antineutrino oscillation studies, for instance accelerator and solar neutrinos compared with reactor antineutrinos. In this sector, we have highlighted the challenge of disentangling the presumed contributions caused by the presence of NSI and the necessity of developing a synergy analysis strategy that involves comparing the results from various experiments. Finally we investigated the possibility of finding decoherence effects in the propagation of neutrino waves packets. This effect can be caused by the non-unitarity of the neutrino propagation and we reviewed the formalism of the Lindebland equation as the ideal framework to set this research.\\
In conclusion, thanks to their wide range of energy and propagation length, neutrinos can play a fundamental role in the search for evidence of QG in the astrophysical sector. Therefore, future experiments may benefit from studying the neutrino physical sector in a multi-messenger approach to investigate potential QG effects and shed light on the real nature of the spacetime.


\vspace{15pt}




\textbf{Acknowledgments:}\\
The Authors would like to acknowledge networking support by the COST Action 18108.
M.D.C.T. would like to thank Giovanni Amelino-Camelia and Francesco Vissani for the useful discussions and suggestions.

\textbf{Conflicts of interest:}\\
The authors declare no conflict of interest.






\end{document}